# On the spin–boson model with a sub–Ohmic bath

Stefan K. Kehrein[1] and Andreas Mielke[2]

Institut für Theoretische Physik,
Ruprecht–Karls–Universität,
Philosophenweg 19,
D-69120 Heidelberg, F.R. Germany

July 30, 2018

## Abstract

We study the spin–boson model with a sub–Ohmic bath using infinitesimal unitary transformations. Contrary to some results reported in the literature we find a zero temperature transition from an untrapped state for small coupling to a trapped state for strong coupling. We obtain an explicit expression for the renormalized level spacing as a function of the bare papameters of the system. Furthermore we show that typical dynamical equilibrium correlation functions exhibit an algebaric decay at zero temperature.

---

[1] E–mail: kehrein@marvin.tphys.uni-heidelberg.de
[2] E–mail: mielke@hybrid.tphys.uni-heidelberg.de

The problem of a small system coupled to environmental degrees of freedom plays an important role in various fields of physics. The spin-boson model is one important example in this class of models. Here a two level system, i.e. a pseudo spin, is coupled to environmental degrees of freedom modeled by non–interacting harmonic oscillators, i.e. bosons. It is defined by the Hamiltonian

$$H = -\frac{1}{2}\Delta_0\sigma_x + \frac{1}{2}\sigma_z \sum_k \lambda_k(b_k^\dagger + b_k) + \sum_k \omega_k b_k^\dagger b_k + E_0 \ . \qquad (1)$$

$\Delta_0$ is the level splitting of the two level system. The couplings $\lambda_k$ and and the bath frequencies $\omega_k$ enter only via the combination $J(\omega) = \sum_k \lambda_k^2 \delta(\omega - \omega_k)$. This model is a paradigm in the theory of dissipative quantum systems. It has a wide range of applications. For instance it has been used to study dissipative tunneling. In this case $\sigma_z$ plays the role of the coordinate of the tunneling particle, $\Delta_0$ is the bare tunneling frequency. We will use this language in this letter. The model has been investigated by various methods. For a review of applications, methods and results we refer to [1, 2].

One usually assumes that the spectral function $J(\omega)$ behaves as $\omega^s$ with $s > 0$ at low frequencies. Furthermore $J(\omega)$ should contain some high frequency cutoff $\omega_c$. The case $s = 1$ is called the Ohmic bath, it has been widely studied and is well understood. As a function of the coupling strength the model shows a zero temperature transition from a trapped to an untrapped state. This has been shown rigorously by Spohn and Dümcke [3]. They mapped the spin–boson model to an Ising model on $\Re$ with a long–range interaction by integrating out the bosonic degrees of freedom. The corresponding Ising model shows a phase transition for $0 < s \leq 1$. The point $s = 1$ is the phase boundary, for $s > 1$ the Ising model never shows long–range order. The ordered phase of the Ising model corresponds to the trapped state in the spin–boson model. Thus in the super–Ohmic case $s > 1$ the particle is never localized, whereas in the Ohmic case one has a transition from an untrapped state for small coupling to a trapped state for sufficiently large coupling.

In the literature about the spin–boson problem exists some confusion about the existence of such a transition for the sub–Ohmic bath $s < 1$ as within some approximation schemes the particle is localized for any nonzero sub–Ohmic coupling at zero temperature [1, 2]. In contrast the corresponding results for the Ising model on $\Re$ indicate a transition as a function of the coupling strength [3]. However, for $s < 1$ these results for the Ising model have not been carried over to the spin–boson model rigorously. Still, based on the findings of Spohn and Dümcke it would appear surprising not to find corresponding transition in the spin–boson model. It should be noted that the sub–Ohmic case is of less physical interest as compared to Ohmic or super–Ohmic coupling, still it contributes to the general understanding of the parameter space of the spin–boson problem. Also one can learn something about possible problems of adiabatic renormalization schemes as we will see.

In Ref. [4] we recently studied the spin–boson model with an Ohmic bath using a new method proposed by Wegner [5]. This method simplifies a given Hamiltonian using infinitesimal unitary transformations. The advantage of infinitesimal unitary transformations is that one is able to eliminate couplings between states with a large energy difference first, couplings between states with smaller energy differences later. In this manner infinitesimal unitary transformations automatically seperate different energy scales. A continuous unitary transformation can be written as a differential equation for the Hamiltonian or equivalently as a set of coupled differential equations for the coupling constants. In the case of a dissipative quantum system, the goal of the transformation is to decouple the system from the bath. This yields an effective Hamiltonian for the system, which can be used to calculate e.g. dynamical quantities. In the case of the

spin–boson model the effective Hamiltonian is only a two–state system that can be described by an energy shift and a renormalized level spacing. The advantage of this method is that the unitarity of the time evolution of system plus bath is not destroyed. One is able to obtain a systematic expansion for the long–time dynamics and the static behaviour of the system.

In the present letter we also study the sub–Ohmic case in some detail using this method. Our results will turn out to be in agreement with the rigorous results in Ref. [3]. In Ref. [4] we derived an equation for the renormalized tunneling frequency $\Delta_\infty$ as a function of the bare tunneling frequency $\Delta_0$ and the spectral function $J(\omega)$. We sketch the derivation of this equation and use it to calculate the renormalized tunneling frequency in the sub–Ohmic case. Whereas $\Delta_\infty$ is a continuous function of the coupling strength for the Ohmic bath, the transition seems to be discontinuous in the sub–Ohmic case. We calculate an explicit expression for the critical value of the bare tunneling frequency for which the transition occurs. Furthermore we apply some general results from Ref. [4] concerning time–dependent equilibrium correlation functions to the sub–Ohmic case.

A continuous unitary transformation of the Hamiltonian can be written as a differential equation for the Hamiltonian

$$\frac{dH(\ell)}{d\ell} = [\eta(\ell), H(\ell)] \tag{2}$$

with the initial Hamiltonian $H(0) = H$ given in Eq. (1). $\eta$ is the generator of the transformation and depends on the flow parameter $\ell$. $\eta$ is an antihermitian operator and we choose

$$\eta = \frac{i}{2}\sigma_y \sum_k \eta_{k,y}(b_k + b_k^\dagger) - \frac{1}{2}\sigma_z \sum_k \eta_{k,z}(b_k - b_k^\dagger) \ . \tag{3}$$

$\eta_{k,y}$ and $\eta_{k,z}$ are $\ell$–dependent parameters. Performing the transformation the Hamiltonian becomes $\ell$–dependent as well. In addition to the coupling of $\sigma_z$ to the bosonic modes, new couplings are generated. One of these new terms is of the form $\sigma_y \sum_k \xi_k(b_k - b_k^\dagger)$. It can be avoided by a suitable choice of the parameters $\eta_{k,y}$ and $\eta_{k,z}$. The other terms describe a coupling of the spin to more than one bosonic mode. In Ref. [4] we showed that within a very good approximation one can restict oneself to a linear coupling of the spin to the bath. Thus one can neglect the normal ordered part of the higher interactions. As a consequence the Hamiltonian has the form given in (1) for any $\ell$, but now the parameters $\lambda_k$ and $\Delta$ depend on $\ell$. Using

$$\eta_{k,y} = -\Delta\lambda_k f(\omega_k, \ell), \quad \eta_{k,z} = \omega_k \lambda_k f(\omega_k, \ell) \tag{4}$$

the $\ell$ dependence of the parameters is given by the flow equations

$$\frac{d\lambda_k}{d\ell} = -\lambda_k \left(\omega_k^2 - \Delta^2\right) f(\omega_k, \ell) \ , \tag{5}$$

$$\frac{d\Delta}{d\ell} = -\Delta \sum_k \lambda_k^2 f(\omega_k, \ell)(2n_k + 1) \ , \tag{6}$$

$$\frac{dE_0}{d\ell} = -\frac{1}{2}\sum_k \omega_k \lambda_k^2 f(\omega_k, \ell) \ , \tag{7}$$

where $n_k = <b_k^\dagger b_k> = (\exp(\beta\omega_k) - 1)^{-1}$. For details of the calculation we refer to Ref. [4]. Introducing the $\ell$–dependent spectral function $J(\omega, \ell) = \sum_k \lambda_k^2(\ell)\delta(\omega - \omega_k)$ the flow equations can be written as

$$\frac{d\Delta}{d\ell} = -\Delta \int d\omega J(\omega, \ell) f(\omega, \ell)(2n(\omega) + 1) \ , \tag{8}$$

$$\frac{\partial J(\omega,\ell)}{\partial \ell} = -2(\omega^2 - \Delta^2)f(\omega,\ell)J(\omega,\ell) , \qquad (9)$$

with the occupation number $n(\omega) = (e^{\beta\omega}-1)^{-1}$. These two equations can be combined into a single equation

$$\frac{d\ln\Delta}{d\ell} = \frac{1}{2}\int d\omega \frac{\partial J(\omega,\ell)}{\partial \ell}\frac{2n(\omega)+1}{\omega^2 - \Delta^2} . \qquad (10)$$

Eq. (10) can be solved approximately by replacing $\Delta(\ell)$ by $\Delta_\infty$ on the right hand side as explained in Refs. [4, 6]. This yields the self–consistency condition

$$\ln\frac{\Delta_\infty}{\Delta_0} = -\frac{1}{2}\int d\omega J(\omega,0)\frac{2n(\omega)+1}{\omega^2 - \Delta_\infty^2}. \qquad (11)$$

The integral on the right hand side contains a singularity, it is to be interpreted as a principal value integral. This equation is applicable for any spectral function $J(\omega,0)$. The only restriction is that the coupling must not be too strong, since then processes involving more than one bosonic mode become important. It can be shown that if such proccesses are included initially in the Hamiltonian, an equation of the form (11) is obtained as well, but with a modified spectral function $J(\omega,0)$. Thus the main problem is not the occurence of multi–boson processes, but the correlations between the bosonic modes which are induced if the coupling is too strong.

A general feature of equation (11) is that slow modes $\omega < \Delta_\infty$ tend to increase the renormalized level splitting whereas fast modes tend to decrease it. This effect can already be observed in a perturbational treatment. The balance of these effects can be essential when one treats a problem with a discrete eigenvalue embedded in a contiuum (compare for example Ref. [6] for the case of the Anderson impurity model). For the spin–boson problem this balance becomes particularly important for the sub–Ohmic bath as will be discussed in the sequel.

In Ref. [4] the function $f(\omega,\ell)$ in Eq. (9) was chosen as

$$f(\omega,\ell) = \frac{\omega - \Delta}{\omega + \Delta}. \qquad (12)$$

With this choice the coupling function $J(\omega,\ell)$ decays monotously as a function of $\ell$ for all values of $\omega$ as depicted in Fig. 1. For the Ohmic bath $J(\omega,0) = 2\alpha\omega\theta(\omega_c - \omega)$ the numerical solution of the differential equations with this choice of $f(\omega,\ell)$ turned out to be in good agreement with the self–consistency equation (11).

For the non–Ohmic bath the coupling is parametrized as

$$J(\omega,0) = K^{1-s}\omega^s\theta(\omega_c - \omega) \qquad (13)$$

with a coupling constant $K$ with the dimension energy. The explicit form of the cutoff function is not important. In contrast to the Ohmic bath, the non-Ohmic bath contains a second energy scale given by the coupling constant $K$. Whereas $\omega_c$ is usually assumed to be much larger than $\Delta_0$, this new energy scale can be of the order of $\Delta_0$. For the super–Ohmic case $s > 1$ and $T = 0$, we can neglect $\Delta$ in the integral on the right hand side of (10) or $\Delta_\infty$ in the integral on the right hand side of (11) for $\Delta_0 \ll \omega_c$ since the integral is not infrared divergent. This yields the well–known result

$$\Delta_\infty = \Delta_0 \exp\left(-\frac{1}{2(s-1)}\frac{\omega_c^{s-1}}{K^{s-1}}\right). \qquad (14)$$

Comparison with a numerical solution of the flow equations shows that this is a very good approximation to the solution of the differential equations for $\Delta_0 \ll \omega_c$. For $\Delta_0 > \omega_c$ the flow equations yield a solution $\Delta(\ell)$ that increases with $\ell$ and $\Delta_\infty$ is larger than $\omega_c$. In this limit the formula (14) is not valid. For the super–Ohmic bath the particle is never localized, there is no transition as in the case of Ohmic dissipation. Our value of $\Delta_\infty$ in (14) is consistent with adiabatic renormalization [1].

We are interested in the sub–Ohmic case. For $s < 1$ $\Delta_\infty$ can be calculated in the limit $\omega_c \to \infty$ since the integral in (11) is not ultraviolet divergent. Then (11) yields

$$\Delta_0 = \Delta_\infty \exp\left(\frac{\pi}{4}\left(\frac{\Delta_\infty}{K}\right)^{s-1} \cot(\frac{\pi}{2}(1-s))\right) \tag{15}$$

If the condition

$$\Delta_0 > \Delta_c \stackrel{\text{def}}{=} K \left(\frac{4\tan(\frac{\pi}{2}(1-s))}{e\,\pi(1-s)}\right)^{\frac{1}{s-1}} \tag{16}$$

is satisfied, the solution for $\Delta_\infty$ is positive and satisfies

$$\Delta_\infty > K \left(\frac{4\tan(\frac{\pi}{2}(1-s))}{\pi(1-s)}\right)^{\frac{1}{s-1}}. \tag{17}$$

Otherwise $\Delta_\infty = 0$. For the sub–Ohmic bath we obtain a transition from an untrapped state for sufficiently large $\Delta_0$ as compared to $K$ to a trapped state for small $\Delta_0$. In contrast to the transition in the Ohmic case, which depends only on the dimensionless coupling strength $\alpha$, the present transition depends on the ratio between $\Delta_0$ and $K$. It occurs for $\Delta_0 = \Delta_c$. Let us remark that $\Delta_c \propto K$ is a simple consequence of the fact that $K$ can be used to fix the energy scale. This has also been obtained by Spohn and Dümcke using a variational calculation [3]. From the self–consistency condition (11) we therefore conclude that the transition is discontinuous since we obtain $\Delta_\infty = \Delta_c \exp(\frac{1}{s-1}) > 0$ for $\Delta_0 = \Delta_c$. As a cautionary remark we would like to emphasize that the nature of the transition is difficult to investigate from the numerical solution of the original differential equation. Discontinuous behaviour is only possible for the asymptotic solution and hence extremely sensitive to accumulation of errors. But away from the transition point we find good agreement with the self–consistency condition and discontinuous behaviour at the transition point is more compatible with the numerics. In any case the existence of a transition from an untrapped to a trapped state for increasing coupling $K$ is certain.

Using (11) it is a straightforward problem to study $\Delta_\infty$ or the critical value $\Delta_c$ as a function of $\omega_c$. The results (15) and (16) are only valid if $\omega_c$ is large compared to all other energy scales. For smaller values of $\omega_c$ our result $\Delta_c \propto K$ no longer holds.

It is also possible to treat the problem for finite temperature $T > 0$. The reasoning is identical to Ref. [4]. For finite temperature Eq. (11) goes over into

$$\Delta_0 = \Delta_\infty \exp\left(\frac{\pi}{4}\left(\frac{\Delta_\infty}{K}\right)^{s-1} \cot(\frac{\pi}{2}(1-s))\right) F\left(\frac{\Delta_\infty}{K}, \frac{T}{K}\right) \tag{18}$$

where

$$F\left(\frac{\Delta_\infty}{K}, \frac{T}{K}\right) = \exp\left(\frac{1}{2}\int d\omega \frac{\omega^s}{K^{s-1}} \frac{\coth(\frac{\beta\omega}{2})-1}{\omega^2 - \Delta_\infty^2}\right). \tag{19}$$

For small values of $T < \Delta_\infty$ one finds $F < 1$. Therefore $\Delta_\infty(T)$ increases with $T$ for small $T$. For large $T$, $F$ becomes large and $\Delta_\infty(T)$ decreases and finally tends to zero for some critical temperatue $T_c$. This behaviour is similar to the Ohmic case [4]. The transition from a finite value of $\Delta_\infty$ at low temperatures to $\Delta_\infty = 0$ at high temperatures can be interpreted as a transition from coherent to completely incoherent motion of the tunneling particle.

Equilibrium correlation functions at $T = 0$ can also be calculated as in Ref. [4]. We consider the autocorrelation function

$$\hat{C}(t) = \frac{1}{2}\langle \sigma_z(t)\sigma_z(0) + \sigma_z(0)\sigma_z(t)\rangle_\beta = \frac{1}{2}\text{Tr}\Big(\rho_{eq}[\sigma_z(t)\sigma_z(0) + \sigma_z(0)\sigma_z(t)]\Big) \;, \tag{20}$$

where

$$\rho_{eq} = \exp(-\beta H)/\text{Tr}\Big(\exp(-\beta H)\Big) \;. \tag{21}$$

Since the Hamiltonian has a simple form for $\ell = \infty$, we want to calculate the expectation values in this limit. The advantage is that for $\ell = \infty$ the equation of motion for operators becomes very simple and can be solved explicitly. Therefore we do not destroy the unitarity of time evolution in contrast to many other approximations. Furthermore it is very easy to perform the thermal average. The price we have to pay is that we have to calculate the operator of interest at $\ell = \infty$. To do this we must solve the flow equation

$$\frac{d\sigma_z(\ell)}{d\ell} = [\eta(\ell), \sigma_z(\ell)] \quad, \quad \sigma_z(0) = \sigma_z \;. \tag{22}$$

Clearly the expression for $\sigma_z(\ell)$ is rather complicated. Fortunately it turns out that one can obtain the correct long–time dynamics with the ansatz

$$\sigma_z(\ell) = h(\ell)\sigma_z + \sigma_x \sum_k \chi_k(\ell)(b_k + b_k^\dagger) \;, \tag{23}$$

i.e. terms containing more than one bosonic operator are neglected. We analyzed the flow equation for $h(\ell)$ and $\chi_k(\ell)$ in [4]. In the following we report the general results obtained in [4] and apply these to the sub–Ohmic bath. The Fourier–transform of $\hat{C}(t) = \frac{1}{\pi}\int_{-\infty}^\infty C(\omega)\, e^{-i\omega t}\, d\omega$ can be written in the form

$$\begin{aligned} C(\omega) &= \frac{\pi h^2(\infty)}{2}\delta(\omega - \Delta_\infty) + \frac{\pi}{2}\left(\int_0^\infty h\Delta f(\omega, \ell)\sqrt{J(\omega,\ell)}d\ell\right)^2 \theta(\omega) + (\omega \leftrightarrow -\omega) \\ &= \frac{\pi h^2(\infty)}{2}\delta(\omega - \Delta_\infty) + \pi\left(\int_0^\infty \frac{h\Delta}{\omega^2 - \Delta^2}\frac{\partial\sqrt{J(\omega,\ell)}}{\partial\ell}d\ell\right)^2 \theta(\omega) + (\omega \leftrightarrow -\omega) \end{aligned} \tag{24}$$

If $\Delta_\infty = 0$ one finds $h(\infty) \neq 0$ as expected. Hence the particle is localized and $\hat{C}(t)$ tends to a constant for large $t$. If $\Delta_\infty > 0$ one has $h(\infty) = 0$ and (24) is correct to leading order in $\omega$, i.e. corrections are at least of order $J(\omega, 0)^2$ for small $\omega$. For $\omega \ll \Delta_\infty$, $J(\omega, \ell)$ tends to zero rapidly. Therefore we can replace $h$ and $\Delta$ in the integral by their initial values. This shows that $C(\omega) \propto J(\omega) \propto \omega^s$ for small $\omega$ and consequently

$$\hat{C}(t) \propto t^{-s-1} \tag{25}$$

for large times. This long–time behaviour of the equilibrium correlation function holds for general $s$ if the particle is not localized. Therefore the asymptotic dynamics is a power–law relaxation in the disordered phase at zero temperature. Let us mention that Monte–Carlo simulations for this correlation function in the sub–Ohmic case reported in [2] indicate an algebraic decay.

As already mentioned in the beginning of this letter, there are various claims in the literature that the particle is always localized in the sub–Ohmic case for zero temperature [1, 2] in contradiction to our results reported here. It is interesting to analyze the reasoning leading to this conclusion. One important method for treating the spin–boson problem is the "non–interacting blip approximation" (NIBA) [1]. In order to apply this scheme, the typical blip width must be much smaller than the typical distance between blips. For the sub–Ohmic bath this is equivalent to $\Delta_0 \ll K$ [1]. Then the NIBA predicts localization, which is obviously in agreement with our condition (16). When $\Delta_0$ and $K$ are of the same order of magnitude, the NIBA cannot be justified *irrespective of the value of* $\omega_c$. Anyway the NIBA is unreliable for studying the long–time behaviour of equilibrium correlation functions as it gives an exponential decay in Eq. (25) instead of an algebraic decay [1]. In this sense we agree with the conclusions of Ref. [8]. For large times and $T = 0$ our method of continuous unitary transformations is a simpler and systematic approximation scheme.

Another tool for analyzing the spin–boson model is adiabatic renormalization. This relies on successively integrating out high frequencies as compared to the tunneling frequency $\Delta$. If one integrates out the frequencies between some lower cutoff $\omega_l \gg \Delta_0$ and $\omega_c$ one obtains a renormalized tunneling frequency $\Delta'(\omega_l)$ in the zeroth–order adiabatic approximation (for details of the calculation see Ref. [1])

$$\Delta'(\omega_l) = \Delta_0 \, \exp\left(-\frac{1}{2} \int_{\omega_l}^{\omega_c} \frac{J(\omega)}{\omega^2} \, d\omega\right) \qquad (26)$$

Choosing $\omega_l = p\,\Delta_0$ with some number $p$ large compared to unity, the above procedure can be iterated: Since $\Delta'(\omega_l) < \Delta_0$ one next integrates out the frequencies between $\omega_l' = p\,\Delta'(\omega_l)$ and $\omega_l$ and so on. This procedure is depicted in Fig. 2. The iteration converges to a finite value of the tunneling frequency if a nonzero solution of

$$\Delta_\infty = \Delta_0 \, \exp\left(-\frac{(K/p)^{1-s}}{2(1-s)} \, \Delta_\infty^{s-1}\right) \qquad (27)$$

exists (we have taken the limit $\omega_c \to \infty$). But this depends on the ratio $K/p$. Above a certain critical value only the trivial solution $\Delta_\infty = 0$ exists. Strictly speaking this renders the adiabatic renormalization scheme useless in the sub–Ohmic case as localization or non–localization cannot depend on the arbitrary number $p$. On the other hand one easily checks that $p$ plays no similar role in the Ohmic or super–Ohmic case.

In contrast in the method employed in this letter the modes are decoupled in a continuous manner as depicted in Fig. 1. Here $1/\sqrt{\ell}$ corresponds to the energy difference that is just being decoupled for a certain value of $\ell$. *The fundamentally important separation of energy scales in the adiabatic renormalization scheme* (or more generally in the "poor man's" scaling approach) *is with respect to the total energy of the modes as measured from some fixed zero point of energy, whereas our procedure separates any two modes with respect to their specific energy difference.* To deal with large energy differences first and then proceeding to smaller energy differences is intuitively the safe order of things, e.g. this is also one of the basic concepts underlying Wilson's numerical renormalization approach [9]. From this viewpoint one can also understand why adiabatic renormalization works for the Ohmic and super–Ohmic case but not for sub–Ohmic coupling: Low energy modes become more and more important for smaller values of $s$ in Eq. (13).

The main price we have to pay is that our flow equations are more complicated than the corresponding renormalization group equations. Renormalization group equations are differential equations with respect to the ultraviolet cutoff and not with respect to energy differences as in our approach. We expect that our scheme can be useful for other models too.

# References

[1] A. J. Leggett, S. Chakravarty, A. T. Dorsey, M. P. A. Fisher, A. Garg and W. Zwerger, Rev. Mod. Phys. **59** (1987) 1; Erratum, Rev. Mod. Phys. **67** (1995) 725.

[2] U. Weiss, Quantum Dissipative Systems (Series in Modern Condensed Matter Physics, Vol. 2, World Scientific, Singapore, 1993).

[3] H. Spohn and R. Dümcke, J. Stat. Phys. **41** (1985) 389.

[4] S. K. Kehrein, A. Mielke and P. Neu, Z. Phys. B **99** (1996) 269.

[5] F. Wegner, Ann. Physik (Leipzig) **3** (1994) 77.

[6] S. K. Kehrein and A. Mielke, Theory of the Anderson impurity model: The Schrieffer–Wolff transformation re–examined (Preprint HD-TVP-95-14, ESI-Preprint 270, cond–mat/9510145, 1995).

[7] P. Lenz and F. Wegner, Flow equations for electron–phonon interactions (Preprint, Univ. Heidelberg, 1996).

[8] S. Chakravarty and J. Rudnick, Phys. Rev. Lett **75** (1995) 501.

[9] K. G. Wilson, Rev. Mod. Phys. **47** (1975) 773.

**Figure captions**

Fig. 1. Decoupling of modes as a function of the flow parameter $\ell$ in our approach, $0 < \ell_1 < \ell_2 < \ell_3$. $J(\Delta_\infty, \ell)$ shows an algebraic decay [4].

Fig. 2. Successive steps of the adiabatic renormalization sceme. The shaded area indicates the modes that are being integrated out in successive steps.

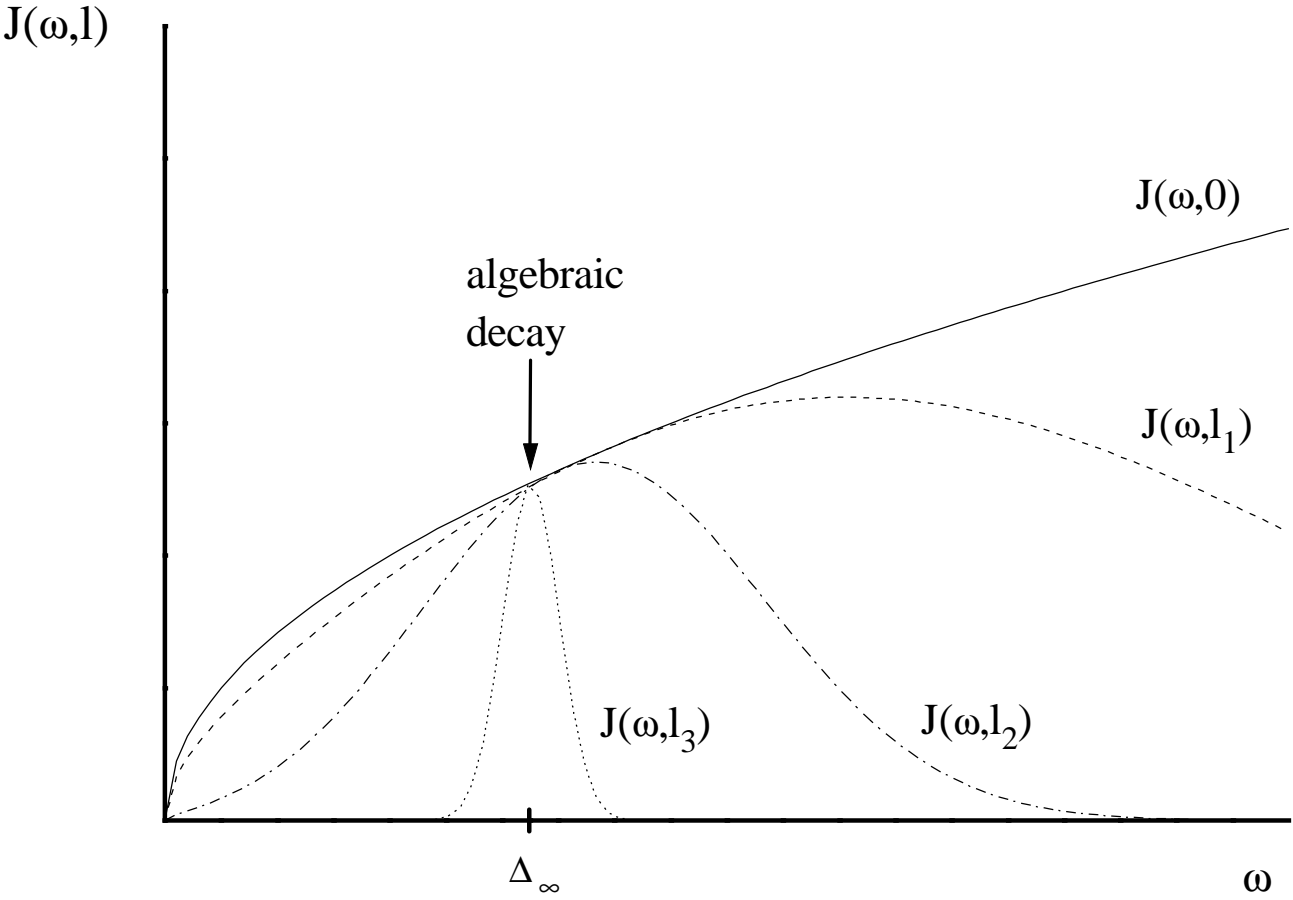

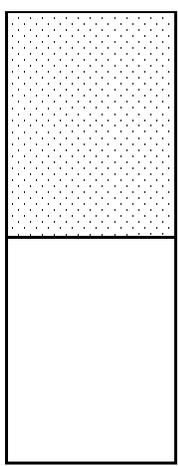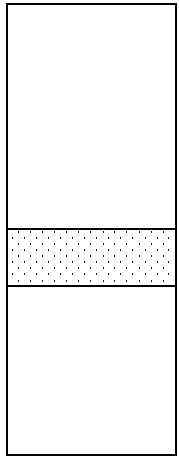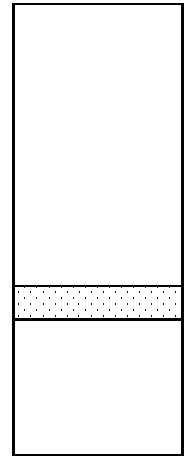